\documentclass[prd,twocolumn,showpacs,floatfix,superscriptaddress]{revtex4}

\usepackage{graphicx}
\usepackage{epsfig}
\usepackage{bm}
\usepackage{amssymb}
\usepackage{float}
\usepackage{amsmath}
\usepackage{dcolumn}
\usepackage{cancel}
\usepackage[colorlinks]{hyperref}
\usepackage[usenames,dvipsnames]{color}
\hypersetup{
     breaklinks=true,
    pdfstartview={FitH},    
    colorlinks=true,       
    linkcolor=blue,          
    citecolor=red,        
    filecolor=magenta,      
    urlcolor=blue,           
    anchorcolor=green,      
    linktocpage=true
}

\def\doi{http://doi.org}
\def\arxiv{http://arxiv.org/abs}

\begin{document}

\title{The generalized second law of thermodynamics with Barrow entropy}

\author{Emmanuel N. Saridakis}
\email{msaridak@phys.uoa.gr}
\affiliation{National Observatory of Athens, Lofos Nymfon, 11852 Athens, 
Greece}
\affiliation{CAS Key Laboratory for Researches in Galaxies and Cosmology, 
Department of Astronomy, University of Science and Technology of China, Hefei, 
Anhui 230026, P.R. China}
\affiliation{School of Astronomy, School of Physical Sciences, 
University of Science and Technology of China, Hefei 230026, P.R. China}

\author{Spyros Basilakos}\email{svasil@academyofathens.gr}
\affiliation{Academy of Athens, Research Center for Astronomy and
Applied Mathematics, Soranou Efesiou 4, 11527, Athens, Greece}

\begin{abstract}
We investigate the validity of the generalized second law of 
thermodynamics,    applying  Barrow entropy for the horizon entropy. 
The former arises from the fact that the 
black-hole surface may be deformed due to quantum-gravitational 
effects, quantified by a new 
exponent $\Delta$.  
 We calculate the entropy time-variation in a universe filled with the matter 
and dark energy fluids, as well as the corresponding quantity for the apparent 
horizon. We show that although in the case $\Delta=0$, which corresponds to 
usual entropy, the  sum of the entropy enclosed by the 
apparent horizon plus the entropy of the   horizon itself is always a 
non-decreasing function of time and thus the generalized second law of 
thermodynamics is valid,  in the case of Barrow entropy this is not true 
anymore, and    the generalized second law of 
thermodynamics may be violated, depending on the universe evolution.
Hence, in order not to have violation, the deformation from standard 
Bekenstein-Hawking expression 
should be
small as expected.

 \end{abstract}


 \maketitle

\section{Introduction}

There is a well known analogy between black-hole physics and thermodynamics. 
 In particular, one can attribute to a black hole a 
specific temperature and entropy, which depend on the black-hole horizon 
\cite{Gibbons:1977mu}. Inspired by this, an extension of this analogy was 
proposed, namely the conjecture of the ``thermodynamics of spacetime'', 
according to which   one can apply thermodynamics in the horizon of the 
universe. In particular,   thermodynamical laws are applied on the horizon 
itself, considered as a system 
separated not by  a diathermic wall but by a causality barrier (i.e. the 
``system'' is composed by the degrees of freedom beyond the horizon)
\cite{Jacobson:1995ab,Padmanabhan:2003gd,Padmanabhan:2009vy}. Thus, these laws 
are interpreted in terms of area of local Rindler horizons
and energy flux \cite{Jacobson:1995ab}, and  heat 
is incorporated  as energy that flows through the causal horizon. Concerning 
the relations for temperature and entropy of the horizon, these are given 
by the  corresponding relations of black hole thermodynamics, but 
with the universe horizon  in place of the black-hole 
horizon.

Applying the first law of thermodynamics on the apparent horizon one can extract 
the Friedmann equations, and reversely one can express the Friedmann equations 
as the the first law \cite{Frolov:2002va,Cai:2005ra,Akbar:2006kj}. This 
procedure proves to be applicable  both in general relativity as well as  in a 
variety of modified gravity theories, despite the fact 
that in the latter 
theories the entropy relation is in general modified
\cite{Akbar:2006er,Paranjape:2006ca,Sheykhi:2007zp,Jamil:2009eb,
Cai:2009ph,
Wang:2009zv,
Jamil:2010di, Gim:2014nba, Fan:2014ala,Lymperis:2018iuz}.  Concerning the 
second law, which in black-hole physics   had been extended to the 
``generalized second law of  thermodynamics'',
namely that the usual entropy plus the black-hole horizon entropy
 is a non-decreasing function of time \cite{Bekenstein:1974ax,Unruh:1982ic}, one 
can also 
apply it to the universe 
horizon  and claim that  the total entropy of the interior of the 
universe plus the entropy of its horizon should be  a non-decreasing 
function of time \cite{Barrow:1988yc}. Note that this statement is proven to be 
always valid for general relativity in  Friedmann-Robertson-Walker geometry, 
however it is not always the case in 
modified theories of gravity and hence one can apply it in order to extract 
constraints on them 
\cite{CANTATA:2021ktz,Wang:1999ni,Setare:2007at,Fouxon:2008pz,Wu:2008ir,
Akbar:2008vz,
Sheykhi:2008qs,Sheykhi:2008qr,Jamil:2010di,Mazumder:2009zza,Karami:2010zz,
Karami:2012fu,Bamba:2012rv,Karami:2014tsa}.   Since the generalized 
second law of thermodynamics is fundamental in physics,  
its (conditional or complete) violation acts 
as a strong argument against the underlying theory.

Recently, Barrow \cite{Barrow:2020tzx}   considered the case that 
quantum-gravitational effects might bring about intricate, fractal structure on 
the black hole surface and thus changing its actual horizon area. This in turn 
led to a new black hole entropy relation, i.e.
\begin{equation}
\label{Barrowentropy}
S_B=
\left (\frac{A}{A_0} \right )^{1+\Delta/2}, 
\end{equation}
where $A$ is the usual horizon area  and $A_0$ the Planck area.
Note that  the above deformed entropy differs from the usual 
``quantum-corrected'' one with logarithmic corrections 
\cite{Kaul:2000kf,Carlip:2000nv}, however
it resembles  Tsallis 
nonextensive entropy
\cite{Tsallis:1987eu,Wilk:1999dr,Tsallis:2012js}, nevertheless the 
involved physical principles and foundations   
are radically different.
The 
quantum-gravitational deformation is quantified by the new 
exponent $\Delta$.  The value $\Delta=0$ 
corresponds to the   simplest horizon structure, and in this case we obtain the 
standard 
Bekenstein-Hawking entropy, while   $\Delta=1$ 
corresponds to   maximal deformation.   

 In the present work we are interested in examining the validity of the 
generalized second law of thermodynamics, but applying the Barrow entropy on 
the horizon   instead of the standard Bekenstein-Hawking one.

\section{Generalized Second Law of Thermodynamics} 

In this section we will apply the generalized second law of thermodynamics in 
the universe using Barrow entropy. We consider 
  a spatially homogeneous and isotropic  Friedmann-Robertson-Walker metric
\begin{equation}
\label{FRWmetric}
ds^{2}=-dt^{2}+a^{2}(t)\delta_{ij}dx^{i}dx^{j}\,.
\end{equation}  
Additionally, we consider that the universe is filled with matter and dark 
energy 
perfect fluids, with energy density and pressure  $\rho_{DE}$, $\rho_{m}$ and 
$p_{DE}$, $p_{m}$ respectively (as usual we do not take into account the 
  black-hole
formation inside the universe and its effect on entropy).
 Thus, the two  Friedmann equations are
\begin{eqnarray}
\label{FRWFR1}
&&H^2=\frac{8\pi G}{3}\left(\rho_m+\rho_{DE}\right)\\
&&\dot{H}=-4\pi G \left(\rho_m+p_m+\rho_{DE}+p_{DE}\right),
\label{FRWFR2}
\end{eqnarray}
with $H=\dot{a}/a$   the Hubble parameter and where a dot denotes the 
derivative with respect to $t$.
Note that the
aforementioned framework holds in general, independently of the specific dark
energy description. Finally, the  conservation of  the total energy-momentum 
tensor gives:
\begin{equation}
 \dot\rho_{DE}+3H(1+w_{DE})\rho_{DE}+ \dot\rho_m+3H(1+w_m)\rho_m=0,
 \label{conserv}
\end{equation}
where $w_i=p_i/\rho_i$ denotes the  equation-of-state parameter of the 
corresponding sector,  which throughout this work are considered 
general and not constant. We mention that the above relation holds both in 
the 
usual case, as well as in the case where the two sectors are allowed to 
mutually 
interact.
 
Concerning the universe horizon, that will be the boundary of the 
thermodynamical system, although in the literature there have been discussed  
various choices, there are many arguments that the appropriate one should be 
the the apparent horizon, which is a
marginally trapped surface with vanishing expansion 
given by \cite{Hayward:1997jp,Hayward:1998ee,Bak:1999hd}:
\begin{equation}
\label{apphor}
 \tilde{r}_A=\frac{1}{\sqrt{H^2+\frac{k}{a^2}}},
\end{equation}
where $k$ quantifies the spatial curvature which is set to zero in the present 
work. Hence, the first Friedmann equation (\ref{FRWFR1}) becomes
\begin{equation}
\label{fr1rev}
\frac{1}{\tilde{r}_A^2}=\frac{8\pi G}{3}(\rho_{DE}+\rho_m).
\end{equation}

We are going to investigate whether the sum of the entropy enclosed by the 
apparent horizon plus the entropy of the apparent horizon itself, is not a
decreasing function of time. We will start by calculating the former and then 
the latter.

In general, the apparent horizon $\tilde{r}_A$ is
time-dependent. Hence, a change $d\tilde {r}_A$ in time interval $dt$ will 
bring about a 
volume-change $dV$, while the energy and entropy of the universe fluids will 
change by $dE$ and $dS$ respectively. Now, the first law of
thermodynamics applied in the universe is written as $TdS=dE+PdV$, and 
therefore the  dark-energy
and  dark-matter entropies read \cite{Wang:2005pk}:
\begin{eqnarray}
\label{en1}
dS_{DE}&=&\frac{1}{T}\Big(P_{DE} dV+dE_{DE}\Big) \\
\label{en2}dS_m&=&\frac{1}{T}\Big(P_m dV+dE_m\Big).
\label{en1b}
\end{eqnarray}
Since the universe volume,  bounded
by the
 apparent horizon, 
 is $V=4 \pi \tilde{r}_A^3/3$, we obtain $ dV=4\pi\tilde{r}_A^2d\tilde
{r}_A$. Additionally, concerning the temperature of the universe fluids, we 
assume it to be the same due to the establishment of equilibrium.
 Dividing (\ref{en1}),(\ref{en1b}) by $dt$ we acquire
\begin{eqnarray}
\label{DS1}
\dot{S}_{DE}&=&\frac{1}{T}\Big(P_{DE}\,
4\pi\tilde{r}_A^2\dot{\tilde
{r}}_A+\dot{E}_{DE}\Big) \\
\dot{S}_m&=&\frac{1}{T}\Big(P_m \, 4\pi\tilde{r}_A^2\dot{\tilde
{r}}_A+\dot{E}_m\Big),
\label{DS2}
\end{eqnarray}
where
\begin{equation}
\dot{\tilde
r}_A=-\dot{H}{\tilde{r}}_A^2,
\label{dotrh}
\end{equation}
as it easily arises differentiating  
(\ref{apphor}).

In order to connect the thermodynamical  quantities 
  $E_i$ and   $P_i$, with the cosmological $\rho_i$
and     $p_i$  ones,  we   straightforwardly use
\begin{eqnarray}
\label{Erelation}
E_{DE}&=&\frac{4\pi}{3}\tilde{r}_A^3\rho_{DE} \\
E_m&=&\frac{4\pi}{3}\tilde{r}_A^3\rho_m.   \label{Erelationb}
\end{eqnarray} 
  Inserting the
time-derivatives of (\ref{Erelation}),(\ref{Erelationb}),
into (\ref{DS1}),(\ref{DS2}) and using (\ref{conserv}), we obtain:
\begin{eqnarray}
\dot{S}_{DE}&=&\frac{1}{T}\left(1+w_{DE} \right)\rho_{DE}\,
4\pi\tilde{r}_A^2\left(\dot{\tilde
{r}}_A-H\tilde{r}_A\right)\label{stotfl1}\\
\dot{S}_m&=&\frac{1}{T}\left(1+w_{m} \right)\rho_{m}\,
4\pi\tilde{r}_A^2\left(\dot{\tilde
{r}}_A-H\tilde{r}_A\right)\label{stotfl2}. 
\end{eqnarray}
These expressions provide the entropy time-variation for the universe interior.

We now proceed to the calculation of  the entropy time-variation for the 
universe horizon. As we mentioned in the introduction, according to   the 
``gravity-thermodynamics'' conjecture the temperature and entropy of the 
horizon will be given by the corresponding quantities of black-hole 
thermodynamics, but with the apparent horizon in place of the black-hole one. 
The horizon temperature   will thus be simply
\cite{Jacobson:1995ab,Cai:2005ra}
\begin{equation}
\label{Threl}
 T_h=\frac{1}{2\pi\tilde{r}_A}.
\end{equation}
Concerning the horizon entropy, the standard choice is to use the 
Bekenstein-Hawking entropy 
  \cite{Jacobson:1995ab,Cai:2005ra} 
$S_h=4\pi\tilde{r}_A^2/(4G)$. However, in the present work we will instead 
apply the Barrow black-hole entropy (\ref{Barrowentropy}), with the 
standard horizon area being $A=4\pi \tilde{r}_A^2$. Therefore, we obtain 
\begin{equation}
\label{Barrowentropy2}
S_h=  \gamma\tilde{r}_A^{\Delta +2},
\end{equation}
with $\gamma\equiv\left(4\pi/A_0\right)^{1+\Delta/2}$.
Finally, a crucial assumption in the 
``gravity-thermodynamics'' conjecture is that  after equilibrium establishes 
the universe fluids acquire the same temperature with the horizon one,  
which is constant or slowly-varying, 
otherwise the energy flow would deform this geometry 
\cite{Izquierdo:2005ku} (in order to avoid applying 
non-equilibrium 
thermodynamics  the assumption of 
equilibrium  is widely used 
\cite{Padmanabhan:2009vy,Frolov:2002va,Cai:2005ra,Akbar:2006kj,Izquierdo:2005ku,
Jamil:2010di,MohseniSadjadi:2005ps,MohseniSadjadi:2006gz,Szydlowski:2018kbk}, 
see also \cite{Mimoso:2016jwg}). Therefore, we can equate $T_h$ in 
(\ref{Threl}) 
with $T$ in 
(\ref{stotfl1}),(\ref{stotfl2}),
  and the temperature is only slowly evolving according to (\ref{Threl}) 
due to 
the slow change of the apparent horizon.  
Lastly, differentiating 
(\ref{Barrowentropy2}) we obtain
\begin{equation}
\label{Shordot}
 \dot{S}_{h}=(\Delta +2)\gamma\tilde{r}_A^{\Delta +1}\,\dot{\tilde {r}}_A.
\end{equation}

We can now calculate the  total entropy time-variation. Adding relations
(\ref{stotfl1}),(\ref{stotfl2}) and (\ref{Shordot}), and replacing $T$ through 
(\ref{Threl}),  we find:
 \begin{eqnarray}
\label{Sdottot2}
 &&
 \!\!\!\!\!\!\!
 \dot{S}_{tot}\equiv\dot{S}_{DE}+\dot{S}_m+ \dot{S}_h\nonumber\\
 &&
 \ \  
 =
8\pi^2\tilde{r}_A^3\left(\dot{\tilde
{r}}_A-H\tilde{r}_A\right)\Big[\!(1\!+\!w_{DE})\rho_{DE}+(1\!+\!w_m)\rho_m\!\Big
]
\nonumber\\
 &&
 \ \ \   \ \  
+(\Delta +2)\gamma\tilde{r}_A^{ \Delta +1}\,\dot{\tilde {r}}_A.
\end{eqnarray}
 Hence, substituting   $\dot{\tilde {r}}_A$ from  (\ref{dotrh}), 
knowing that ${\tilde {r}}_A=H^{-1}$,  and using the two Friedmann equations 
(\ref{FRWFR1}),(\ref{FRWFR2}), we easily find
 \begin{equation}
\label{Sdottotal}
 \dot{S}_{tot}=\frac{2\pi}{G}H^{-5}\dot{H}\left\{
 \dot{H}+ H^2\left[1-\frac{\gamma G}{2 \pi}(\Delta+2)H^{-\Delta}\right]
\right\}.
\end{equation}

Let us now examine the sign of  (\ref{Sdottotal}). In the case where the Barrow 
exponent takes its standard value $\Delta=0$, i.e. in the case of usual 
black-hole thermodynamics, we have 
 \begin{equation}
\label{Sdottotal2}
 \dot{S}_{tot}|_{_{\Delta=0}}=\frac{2\pi}{G}H^{-5}\dot{H}^2\geq0,
\end{equation}
since in this case $\gamma\equiv\left(4\pi/A_0\right)=\pi/G$ (in units where 
$\hbar=k_B = c = 1$). As expected,  $\dot{S}_{tot}|_{_{\Delta=0}}\geq0$ and 
thus the generalized second law of thermodynamic is always valid in a universe 
filled with matter and dark energy sectors and governed by general relativity 
(note that the limiting result  $\dot{S}_{tot}|_{_{\Delta=0}}=0$ is obtained 
for the de Sitter universe).

However, interestingly enough, in the case where the Barrow exponent is 
non-zero, i.e. the quantum-gravitational effects on the entropy switch on, the 
total entropy is not necessarily a non-decreasing function of time, and hence 
the generalized second law can be violated. The   reason is the 
following.

In the case of standard Bekenstein-Hawking universe thermodynamics, 
 $\dot{S}_{h}$  has always the necessary value in order to bring  $ 
\dot{S}_{tot}$ to non-negative values. In particular, for $\dot{H}<0$, 
i.e. in the case where the universe fluids satisfy the null energy condition in 
total, we have 
$\dot{S}_{h}>0$ and thus it leads to $\dot{S}_{tot}\geq0$ even if the universe 
fluids have $\dot{S}_{DE}+\dot{S}_m<0$, while in the case $\dot{H}>0$, i.e for 
the violation of the total null energy condition, where 
$\dot{S}_{h}<0$ we have $\dot{S}_{DE}+\dot{S}_m>0$  always in a sufficient 
amount to make  $\dot{S}_{tot}\geq0$.
However, in the case where Barrow entropy is used the horizon contributes with 
a changed  $\dot{S}_{h}$, which does not satisfy the aforementioned conditions 
anymore (in the case where it is positive it is not always sufficiently 
positive to counterbalance  cases where $\dot{S}_{DE}+\dot{S}_m<0$, while in 
the case where it is negative it is not always sufficiently close to zero 
in order to be counterbalanced  by the $\dot{S}_{DE}+\dot{S}_m>0$).
Therefore, the use of the deformed, Barrow entropy, leads to the conditional 
violation of the generalized second law of 
thermodynamics, depending on the universe evolution.

 In order to see the above qualitative behavior more transparently we 
proceed to a  quantitative 
investigation, by exploring $ \dot{S}_{tot}$ for different forms of Hubble 
function evolution. For convenience we   use the 
redshift  $  1+z=a_0/a$ (with $a_0=1$ the present scale factor) as the 
independent variable, and thus 
relation  (\ref{Sdottotal}) becomes  
 \begin{eqnarray}
\label{Sdottotalbb}
 &&
 \!\!\!\!\!\!\!\!\! \!\!\!\!\!\!\!\!
 \dot{S}_{tot}(z)=\frac{2\pi}{G}H(z)^{-3}H'(z)(1+z)\Big\{
H'(z)(1+z)
\nonumber\\
&&
\ \ \  \ \ \ \ \ \ \ \ \ \,
- H(z)\Big[1-\frac{\gamma G}{2 
\pi}(\Delta+2)H(z)^{-\Delta}\Big]
\Big\},
\end{eqnarray}
where primes denote derivatives with respect to $z$ and we have used the 
relation $\dot{f}=-(1+z)Hf'$.
 \begin{figure}[ht]
  \hspace{-1.cm}
\includegraphics[scale=0.46]{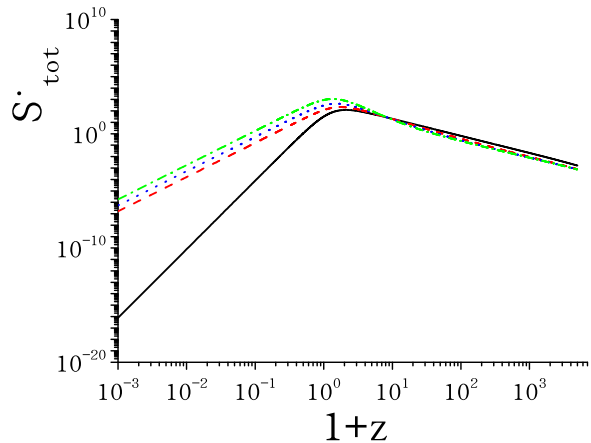}
\caption{
{\it{The evolution of the    entropy time-variation  as a function of the 
redshift $z$ given by  (\ref{Sdottotalbb}), in the case of $\Lambda$CDM 
background evolution (\ref{HLCDM}), for various values of the Barrow exponent 
$\Delta$:  $\Delta=0$ (black - solid), $\Delta=0.3$ (red - dashed), 
$\Delta=0.6$ (blue - dotted), $\Delta=1$ (green - dashed-dotted).
 We have imposed 
 $\Omega_{m0}=0.30$  and  $\Omega_{r0}=10^{-4}$, and we have used units where 
$\hbar=k_B = c = 1=H_0=1$.
}} }
\label{LCDM2}
\end{figure}

As a first example we consider the case where the Hubble 
function evolves as in $\Lambda$CDM cosmology, namely as
  \begin{equation}
H_{\Lambda \text{CDM}}(z) = H_0 \sqrt{\Omega_{m0} (1+z)^3+\Omega_{r0} 
(1+z)^4+\Omega_{\Lambda0} },
\label{HLCDM}
\end{equation}
where $H_0$ is the current Hubble parameter, $\Omega_{i0}$ is the value of the 
density parameter  $\Omega_i=8\pi G\rho_i/(3H^2)$ at present, 
$\Omega_{\Lambda0}=1-\Omega_{m0}-\Omega_{r0}$, and with the subscripts ``m'' and 
``r'' 
denoting the matter and radiation sectors respectively. We insert 
(\ref{HLCDM}) into (\ref{Sdottotalbb}) and in Fig.   \ref{LCDM2} we depict 
the evolution of 
$\dot{S}_{tot}(z)$ for various values of the Barrow exponent $\Delta$.
 As we observe, $\dot{S}_{tot}$ is always positive, throughout cosmological 
evolution, and for all values of $\Delta$. This is an advantage for Barrow 
entropy, since it implies that for $\Lambda$CDM background the generalized 
second law of thermodynamics is always valid.

\begin{figure}[ht]
  \hspace{-1.cm}
 \includegraphics[scale=0.46]{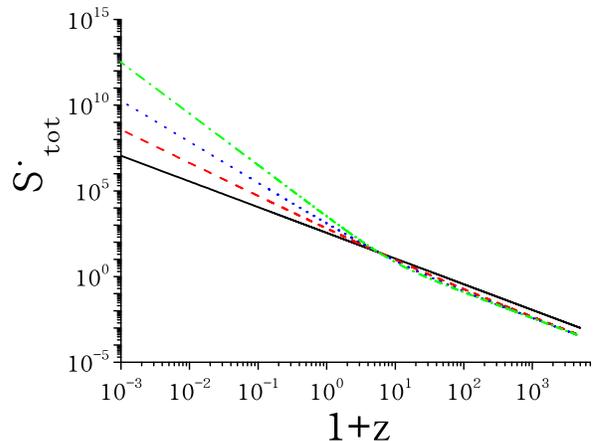} \\
   \hspace{-1.cm}
 \includegraphics[scale=0.46]{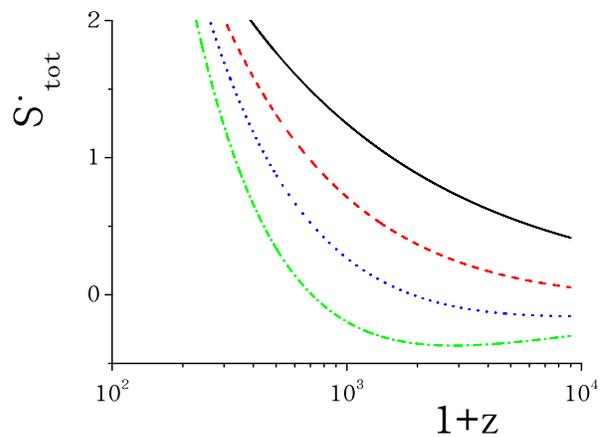} 
\caption{
The evolution of the    entropy time-variation  as a function of the 
redshift $z$ given by  (\ref{Sdottotalbb}), in the case of power-law
background evolution (\ref{Hpower}), for various values of the 
Barrow exponent $\Delta$.
\it{Upper panel: $n=2/3$ and  $\Delta=0$ (black - solid), $\Delta=0.3$ (red - 
dashed), 
$\Delta=0.6$ (blue - dotted), $\Delta=1$ (green - dashed-dotted).
Lower 
panel (mind the difference in the axes): $n=2$ and   $\Delta=0$ (black - 
solid), $\Delta=0.3$ (red - dashed), 
$\Delta=0.6$ (blue - dotted), $\Delta=1$ (green - dashed-dotted).
We have imposed 
 $\Omega_{m0}=0.30$  and  $\Omega_{r0}=10^{-4}$, and we have used units where 
$\hbar=k_B = c = 1=H_0=1$.
}}
\label{power12}
\end{figure}

Let us proceed by investigating a general power-law cosmological evolution of 
the form $a(t)=a_1 t^n$, which implies that 
  \begin{equation}
H_{power}(z) = H_0   (1+z)^{1/n} 
\label{Hpower}
\end{equation}
with $H_0=n a_1^{1/n}$, which is a well studied scenario in cosmology.  
Inserting 
(\ref{Hpower}) into (\ref{Sdottotalbb}) 
we find
  \begin{eqnarray}
 &&
  \!\!\!\!\!\!\!\!\! 
\!\!\!\!\!\!\!\!
\!\!\! 
\dot{S}_{tot}(z)=\frac{\left[H_0(z+1)^{1/n}\right]^{-1-\Delta}}{ 
Gn^2} \,\Big\{
 Gn\gamma(2+\Delta)
 \nonumber\\
 &&
 \ \ \  \ \ \ \ \ \ \ \ \ \,\ \ \ \,
 +2(1-n)\pi \left[H_0(z+1)^{1/n}\right]^{\Delta}
 \Big\},
 \label{powersol1}
\end{eqnarray}
and in Fig.   \ref{power12} we depict 
the evolution of 
$\dot{S}_{tot}(z)$ for various values of the Barrow exponent $\Delta$, 
considering $n=2/3$ (upper panel) and $n=2$ (lower panel).
As we can see, in the case where $n<1$  the    entropy time-variation is 
always positive for all $\Delta$ values, and thus the generalized 
second law of thermodynamics is always valid (this can be immediately seen from 
the analytical expression (\ref{powersol1}) too). However, in the case $n>1$ we 
observe that for suitably large $\Delta$ values  $\dot{S}_{tot}$ can become 
negative in the past  (this can be implied from 
 (\ref{powersol1}) too), and thus the generalized 
second law is violated. Hence, these parameter regions should be avoided.

 We close this section by mentioning that although Barrow entropy accounts 
for quantum-gravitational phenomena, since the gravity-thermodynamics 
conjecture incorporates holography, which in turn dually connects the 
very small 
with the very large,   one in principle may have the reflection of these 
phenomena on the classical universe horizon itself (this is also the case of 
the usual logarithmic-corrected entropy due to quantum effects, which is widely 
studied in the universe-thermodynamics framework as well 
\cite{Kaul:2000kf,Carlip:2000nv,Cai:2008ys,Radicella:2010ss}).
Nevertheless, what  is the
value of $\Delta$ in the real world is a different thing, and actually one 
expects it to be close to the standard value $\Delta=0$. Indeed, recent
cosmological observational constraints  on       $\Delta$ reveal that 
the value zero is indeed inside the 1$\sigma$ region 
\cite{Anagnostopoulos:2020ctz,Barrow:2020kug}.

\section{Conclusions}\label{conclusions}

 In this work we investigated the validity of the generalized second law of 
thermodynamics,  but using for the horizon entropy the 
 Barrow one. Specifically, Barrow entropy arises from the fact that the 
black-hole surface may be deformed due to quantum-gravitational 
effects, and its deviation from Bekenstein-Hawking one is quantified by a new 
exponent $\Delta$.
 
 We calculated the entropy time-variation in a universe filled with the matter 
and dark energy fluids, as well as the corresponding quantity for the apparent 
horizon. As we showed, although in the case $\Delta=0$, which corresponds to 
usual entropy, the  sum of the entropy enclosed by the 
apparent horizon plus the entropy of the   horizon itself is always a 
non-decreasing function of time and thus the generalized second law of 
thermodynamics is valid, in the case where quantum-gravitational corrections 
switch on this is not true anymore. Hence, the generalized second law of 
thermodynamics may be conditionally violated, depending on the universe 
evolution.   In particular, in the case of  $\Lambda$CDM background 
cosmological evolution the    entropy time-variation is always positive and 
thus   the generalized second law of 
thermodynamics  is always valid. However, in the case of  a general 
power-law cosmological evolution we found that for power-law exponents larger 
than one and for suitably large $\Delta$ values  the    entropy time-variation 
can be negative in the past, leading to the violation of the generalized second 
law.   Hence, the involved parameter regions should be 
avoided, and therefore if  Barrow 
entropy is the case in nature then its allowed region would be constrained in a
narrow window close to the standard entropy. This reveals the usefulness 
of examining the generalized second 
law of thermodynamics.  

It would be interesting to investigate whether the above result 
remains valid in the case of various gravitational modifications 
instead of general relativity, and whether the known violations of the 
generalized second law in modified gravity may be eliminated using Barrow 
entropy. These studies lie beyond the scope of the present work and are left 
for future projects.

\end{document}